\documentclass[twocolumn,showpacs,preprintnumbers,amsmath,amssymb,prl,aps]{revtex4-1}

\usepackage{epsfig}
\usepackage{graphicx}
\usepackage{bm}
 \usepackage{float} 
\usepackage{amsmath}
\usepackage{gensymb}
\usepackage[colorlinks=true,linkcolor=red,citecolor=blue]{hyperref}%

\begin{document}

\title{Hidden phase in parent Fe-pnictide superconductors}

\author{Khadiza Ali$^1$, Ganesh Adhikary$^2$, Sangeeta Thakur$^3$, Swapnil Patil$^{1,4}$,
Sanjoy K. Mahatha$^5$, A. Thamizhavel$^1$, Giovanni De Ninno$^{2,3}$, Paolo Moras$^5$, Polina M. Sheverdyaeva$^5$, Carlo Carbone$^5$, Luca Petaccia$^3$ and Kalobaran Maiti$^1$}
\altaffiliation{Corresponding author: kbmaiti@tifr.res.in}

\affiliation{$^1$ Department of Condensed Matter Physics and
Materials Science, Tata Institute of Fundamental Research, Homi
Bhabha Road, Colaba,
Mumbai - 400 005, INDIA.\\
$^2$ Laboratory of Quantum Optics, University of Nova Gorica, Nova
Gorica, Slovenia.\\
$^3$ Elettra Sincrotrone Trieste, Strada Statale 14 Km 163.5, I-34149 Trieste, Italy.\\
$^4$ Department of Physics, Indian Institute of Technology (Banaras Hindu University), Varanasi-221005, India\\
$^5$ Istituto di Struttura della Materia, Consiglio Nazionale delle
Ricerche, Trieste, Italy.
}

\date{\today}

\begin{abstract}
We investigate the origin of exoticity in Fe-based systems via studying the Fermiology of CaFe$_2$As$_2$ employing Angle Resolved Photoemission spectroscopy (ARPES). While the Fermi surfaces (FSs) at 200 K and 31 K are observed to exhibit two dimensional (2D) and three dimensional (3D) topology, respectively, the FSs at intermediate temperatures reveal emergence of the 3D topology at much lower temperature than the structural \& magnetic phase transition temperature (170 K, for the sample under scrutiny). This leads to the conclusion that the evolution of FS topology is not directly driven by the structural transition. In addition, we discover the existence in ambient conditions of energy bands related to the cT phase. These bands are distinctly resolved in the high-photon energy spectra exhibiting strong Fe 3$d$ character. They gradually move to higher binding energies due to thermal compression with cooling, leading to the emergence of 3D topology in the Fermi surface. These results reveal the so-far hidden existence of a cT phase in ambient conditions, which is argued to lead to quantum fluctuations responsible for the exotic electronic properties in Fe-pnictide superconductors.
\end{abstract}


\maketitle


The parent compounds of Fe-based superconductors are paramagnetic metals and undergo structural \& magnetic transitions exhibiting spin density wave (SDW) state as the magnetic ground state \cite{review1} with Fe atoms possessing magnetic moment close to a Bohr magneton \cite{fesc2_CaFe2As2,mag_CaFe2As2,KBM-Pramana,mag_BaFe2As2,mag_LaoFeAs}. These materials exhibit varied unusual phenomena involving competing interactions related to magnetic order, superconductivity \cite{coexist_theory,ganesh-EuFe2As2}, etc. Superconductivity in the Fe-based compounds is believed to appear due to spin fluctuations. Recent studies, however, revealed mysterious superconductivity in pressure induced non-magnetic phase \cite{pressure_FeSe}.

The finding of superconductivity under pressure raises concern over the applicability of spin-fluctuation theory of superconductivity \cite{spinfluc_wang,spinfluc_allan}. CaFe$_2$As$_2$ is an archetypical test case for the investigation of such puzzles. It has paramagnetic tetragonal structure at room temperature and undergoes a transition to the orthorhombic antiferromagnetic (AFM) phase below 170 K \cite{fesc2_CaFe2As2,mag_CaFe2As2,KBM-Pramana}. Application of small pressure ($>$~0.35 GPa) collapses the system in its tetragonal symmetry; this is called \textit{collapsed tetragonal} (cT) phase and does not exhibit magnetic order \cite{cT_nonmag1neutron}. Extensive studies have been carried out on this system resulting into conflicting conclusions \cite{pressure_Reviewcanfield2009structural,ganesh_CaFe2As2}. Some studies find $T_c$ as high as 45 K \cite{sc_CaFe2AS2_45k_La} in doped CaFe$_2$As$_2$ under pressure and attributed the superconductivity to cT phase \cite{pressure_CaFe2AS2b,pressure_CaFe2As2_park}. Some other studies do not find superconductivity in cT phase, which is interpreted as a support of the spin-fluctuation theory of superconductivity \cite{presuure_CaFe2As2_yu}. Evidently, the link between superconductivity, structural phase and spin fluctuations is an outstanding issue.  Here, we study the Fermiology of CaFe$_2$As$_2$ at different temperatures and discover evidence of cT phase hidden within the structural phase at atmospheric pressure, which plays important role in deriving puzzling exoticity of these materials.


Single crystals of CaFe$_2$As$_2$ were grown using high temperature solution growth method as described in the Refs. \cite{neraj,mittal}. The composition and structure were verified by energy dispersive analysis of $x$-rays (EDAX) followed by $x$-ray photoemission and $x$-ray Diffraction (XRD). Single crystallinity was ensured by sharp Laue Pattern. It is found that the electronic properties of Fe-based compounds often depend on the sample preparation conditions. For example, SrFe$_2$As$_2$ shows coexisting superconductivity and ferromagnetism; the volume fraction of each phase depends on the preparation conditions i.e. the extent of strain present in the system \cite{SrFe2As2_1}. Therefore, we studied different pieces of single crystalline CaFe$_2$As$_2$ samples and find that they exhibit magnetic \& structural transitions at 170 K for all the samples. Angle resolved photoemission (ARPES) measurements were carried out at Elettra, Trieste, Italy and TIFR, Mumbai using Scienta R4000 WAL electron analyzer with an energy resolution fixed at 15 meV and angle resolution of ${\approx}$ 0.3$^{\circ}$. The samples were cleaved \textit{in situ} along $ab$ plane yielding a mirror like clean surface. The base pressure of the spectrometer chamber was maintained at 4$\times$10$^{-11}$ Torr during cleaving and the photoemission measurements. ARPES is a highly versatile and powerful tool to probe the electronic structure directly. The technique is based on photoelectric effect, which is a fast process enabling sudden approximation to capture the essential features of the experimental results. Thus, ARPES is able to capture local electronic structure at a fast time scale, and is able to reveal the signature of phases hidden to other experimental techniques involving slower time scales.

The electronic structure calculations were carried out using full potential linearized augmented plane wave method (FLAPW) as captured in the Wien2k software \cite{wien2k}. In this self consistent method, the convergence to the ground state was achieved by fixing the energy convergence criteria to 0.0001 Rydberg ($\sim$1 meV) using 10$\times$10$\times$10 $k$-points in the Brillouin zone and for Fermi surface calculations 39$\times$39$\times$10 $k$-points were used. We have used the Perdew-Burke-Ernzerhof generalized gradient approximation (GGA) for the density functional theoretical calculation. The Fermi surfaces were calculated using Xcrysden \cite{crysden}. The collapsed tetragonal phase possesses the same space group but a reduced $c$ axis and slightly increased $a$ axis, which leads to an overall reduction in cell volume. In the tetragonal phase (space group $I4/mmm$), the lattice parameters are, $a$ = 3.8915 \AA, $c$ = 11.69 \AA, and $z_{As}$ = 0.372. Orthorhombic structure of CaFe$_2$As$_2$ appearing at low temperature and ambient pressure has $Fmmm$ space group with the lattice parameters $a$ = 5.506 \AA, $b$ = 5.450 \AA, $c$ = 11.664 \AA. The lattice parameters for the cT phase at $P$ = 0.63 GPa are $a$ = 3.978 \AA, $c$ = 10.6073 \AA; we used $z_{As}$ = 0.372 and 0.3663 for the calculations of cT phase.


\begin{figure}
\includegraphics [scale=1.5, angle=0]{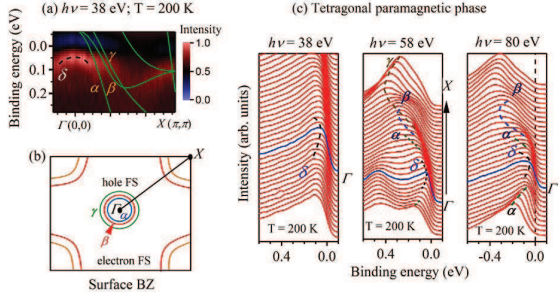}
\caption{ARPES results of CaFe$_2$As$_2$. (a) Second derivative of ARPES data at 200 K ($h\nu$ = 38 eV) exhibiting three energy bands ($\alpha$, $\beta$ and $\gamma$). The lines represent the {\it ab initio} results for the tetragonal structure after 40\% compression of the energy scale. There exists one additional filled band denoted by $\delta$ around $\Gamma$-point. The two energy bands crossing near $X$ point form electron pockets.  (b) Schematic of the Brillouin zone exhibiting three hole pockets around $\Gamma$ corresponding to $\alpha$, $\beta$ \& $\gamma$ bands, and electron pockets at the zone corners. (c) Energy distribution curves (EDCs) at 200 K with $h\nu$ = 38 eV, 58 eV and 80 eV exhibiting distinct signature of the $\delta$ band in the paramagnetic tetragonal phase.}
\vspace{-2ex}
\label{revFig1_SDW}
\end{figure}

We show the ARPES data of CaFe$_2$As$_2$ in Fig. \ref{revFig1_SDW} exhibiting several energy bands and interesting evolutions. In Fig. \ref{revFig1_SDW}(a), the ARPES data collected at 200 K (paramagnetic tetragonal phase) are superimposed with the calculated energy bands obtained by {\em ab initio} density functional theory. The energy scale of the theoretical results is compressed by 40\% to match the experimental bandwidth; such narrowing of the LDA bandwidth found to occur due to electron correlation induced effects \cite{Na-bndstr,DD-PRL} and is the signature of a correlated behavior of Fe 3$d$ electrons in CaFe$_2$As$_2$. Three energy bands denoted by $\alpha$, $\beta$ \& $\gamma$ cross the Fermi level, $\epsilon_F$ forming three hole pockets around the $\Gamma$-point \cite{ARPES_CaFe2As2,Arpes_kondo,Arpes_wang}. Two energy bands crossing $\epsilon_F$ near $X$ point form electron pockets. In addition, there is a {\it mysterious} filled energy band just below $\epsilon_F$ at $\Gamma$, denoted by $\delta$ in the figure, which does not have a counterpart in theoretical results. Distinct signature of the $\delta$-band is observed in Fig. \ref{revFig1_SDW}(c), where energy distribution curves (EDCs) at 200 K with photon energy ($h\nu$) of 38 eV, 58 eV and 80 eV are shown. The signature of the $\delta$ band becomes well defined in the spectra obtained using higher photon energies.

\begin{figure}
\includegraphics [scale=1.5, angle=0]{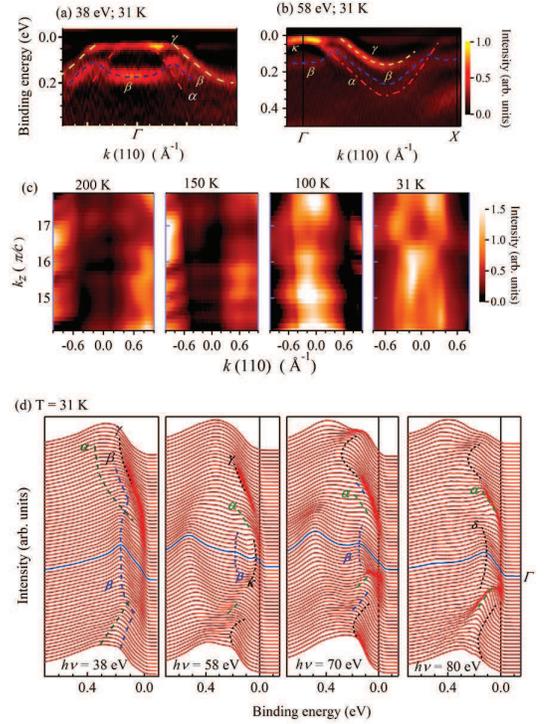}
\caption{Second derivative of the ARPES data collected at 31 K for (a) $h\nu$ = 38 eV and (b) $h\nu$ = 58 eV. The results show distinct signature of folded $\beta$ band due to SDW transition and two hole pockets around $\Gamma$. (c) Fermi surface in the $k_x-k_z$ plane at 200 K, 150 K, 100 K and 31 K. The Fermi surface topology is two dimensional at 200 K and 150 K. A $k_z$-dependence of the Fermi surface is observed at 100 K and 31 K. (d) EDCs at 31 K at $h\nu$ = 38 eV ($\sim 14\pi/c$), 58 eV ($\sim 16\pi/c$), 70 eV ($\sim 17\pi/c$) and 80 eV($\sim 18\pi/c$). $\alpha$ and $\gamma$ bands appear to cross $\epsilon_F$ at all $k_z$ values. EDC collected at 58 eV exhibit an additional band below $\epsilon_F$ denoted by $\kappa$. Distinct $\delta$-band is seen in 80 eV spectra.}
\label{revFig2_kz}
\hspace{-6ex}
\end{figure}

The electronic structure of the magnetically ordered orthorhombic phase is discussed based on the spectra collected at 31 K. The spectra in Figs. \ref{revFig2_kz}(a) and \ref{revFig2_kz}(b) correspond to photon energies of 38 eV and 58 eV, which are close to $k_z$ = 14$\pi/c$ and 16$\pi/c$, respectively; the inner potential, $V_0$ is found to be close to 16 eV. The spectra in both cases exhibit signatures of two hole pockets corresponding to the $\alpha$ and $\gamma$ bands. The $\beta$ band is folded (dashed line for guidance) due to the supercell symmetry formed by the AFM order \cite{ARPES_CaFe2As2,afm4_zabolotnyy2009pi}. Such band folding is also evident in Fig. \ref{revFig2_kz}(b); this photon energy corresponds to the $\Gamma$ point on the $k_z$ axis. The nesting of the $\beta$-band Fermi surface constituted by Fe (3$d_{xz}$, 3$d_{yz}$) states and its eventual folding in the magnetically ordered phase is in line with the expectation as the Fe-layers are sandwiched by As layers, and the hybridized Fe 3$d$ - As 4$p$ states mediate the magnetic exchange coupling between the Fe 3$d$ local moments. The $\alpha$ and $\gamma$ bands cross $\epsilon_F$ in {\it both} the tetragonal and orthorhombic phases, which is consistent with the results obtained from the band structure calculations \cite{ctdft_khadiza}.

The evolution of the Fermiology with temperature depicted in Fig. \ref{revFig2_kz}(c) is surprising. At 200 K, the Fermi surface (FS) is essentially 2D as expected for the tetragonal structure; this 2D topology survives even at 150 K although the system has undergone a first order structural transition to the orthorhombic phase at 170 K. The FS at 150 K is slightly shrinked compared to the FS at 200 K. This weak effect on the Fermi surface across the structural transition is understandable as the lattice constants in the orthorhombic structure are very close to the ones in the tetragonal phase (the overall change is $< \pm$1\%). The calculated Fermi surfaces for $both$ the tetragonal and orthorhombic phases \cite{ctdft_khadiza} corroborate well with the experimental scenario observed here.

Significant change in FS topology emerges at 100 K and becomes clearly visible at 31 K. In Fig. \ref{revFig2_kz}(d), we show the EDCs at 31 K at $h\nu$ = 38 eV, 58 eV, 70 eV, and 80 eV, respectively. Distinct signature of the folded $\beta$ band appears at about 180 meV at $\Gamma$. The data at $h\nu$ = 80 eV exhibit an intense energy band around 120 meV akin to $\delta$ band observed in Fig. \ref{revFig1_SDW} dispersing in opposite direction of the $\beta$ band. The band structure at all the $k_z$-values shown in Fig. \ref{revFig2_kz}(d) exhibits two hole pockets around $k_z$-axis. One distinct band, named $\kappa$, appears below $\epsilon_F$ at $k_z$ (orthorhombic phase) $\approx$ 16$\pi/c$ ($h\nu$ = 58 eV), while is absent at other $k_z$ data shown in the figure. This scenario was interpreted as a transition from 2D to 3D topology of the Fermi surface in the ground state \cite{ARPES_CaFe2As2}.

If 3D topology of the Fermi surface were associated to the orthorhombic structure, the signature of the changed Fermi surface should have appeared soon after
the structural transition occurring at 170 K. Experimental results exhibit a different scenario. Moreover, we discover that the emergence of the changed topology depends on sample although the structural transition occurs at the same temperature in all the cases. Thus, it appears clear that the three dimensional character of the Fermi surface at low temperatures and the structural changes at 170 K are two different phenomena. The features near $\epsilon_F$ exhibit intense signature of folded $\beta$ band at low photon energy spectra ($h\nu$ = 38 eV) and intense $\delta$-band at higher photon energies ($h\nu$ = 80 eV). The comparison of photoemission cross section of various electronic states indicates that $\beta$ band possesses significant As $p$ contributions; this is expected as the magnetic long-range order occurs via intersite exchange coupling involving Fe 3$d$ - As 4$p$ hybridized states. The $\delta$ band appears to have predominantly Fe 3$d$ character.

\begin{figure}
\includegraphics [scale=1.5, angle=0]{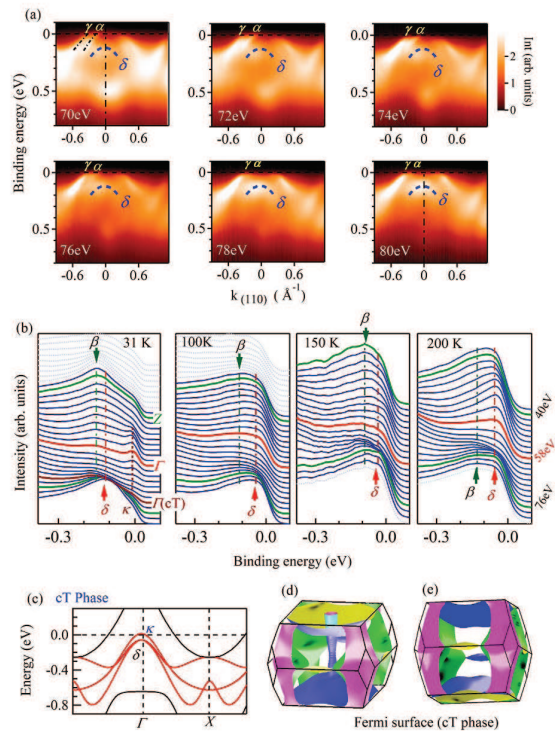}
\caption{(a) Energy bands at 31 K probed with different photon energies exhibiting distinct signature of $\alpha$ and $\gamma$ bands crossing the Fermi level, and one additional band, $\delta$. (b) Normal emission EDCs at 31 K, 100 K, 150 K and 200 K as a function of photon energies. The red and green lines in each plot corresponds to the $\Gamma$ and $Z$-point, respectively in the orthorhombic phase. Maroon line represents the $\Gamma$ point in cT phase at $P$ = 0.63 GPa (denoted by $\Gamma(cT)$. Signatures of $\delta$ and $\beta$ bands are shown by vertical dashed lines. (c) Calculated energy band dispersions of CaFe$_2$As$_2$ in the cT phase with $z_{As}$ = 0.372 (the value in tetragonal phase). Calculated Fermi surfaces in the cT phase with (d) $z_{As}$ = 0.372 and (e)  $z_{As}$ = 0.3663 (experimental value at $P$ = 0.63 GPa).}
\label{Fig3_cT}
\end{figure}

In Fig. \ref{Fig3_cT}, we show the energy bands obtained at 31 K using high photon energies, where the signature of the $\delta$ band is prominent. It is to note here that $\Gamma$ in the orthorhombic and tetragonal phases in abient conditions is represented by the photon energy $\sim$58 eV, while $\Gamma$ in cT phase will correspond to higher photon energy due to compression of $c$-axis. For example, $\Gamma$ corresponds to $h\nu \sim$ 72 eV for the structure at $P$ = 0.63 GPa ($c$ = 10.607 \AA). While $\alpha$, $\beta$ and $\gamma$ bands exhibit evolution in agreement to the theory, intense $\delta$ band is also seen below $\epsilon_F$ with the top of the band at 120 meV binding energy. This band does not shift with the change in photon energy indicating its effective two-dimensional nature. The intensity of the feature is high at high photon energies and gradually reduces when the probing photon energy is decreased.

In Fig. \ref{Fig3_cT}(b), we show the EDCs corresponding to (0,0,$k_z$) points obtained from normal emission ARPES data at different photon energies. Distinct signature of the $\beta$ band is seen at 31 K. The energy band denoted by $\kappa$ (see dashed maroon line in the figure) appears below $\epsilon_F$ near $\Gamma$ on the $k_z$ axis, indicating absence of corresponding Fermi surface as also seen in Fig. \ref{revFig2_kz}(d) ($k_z$ = 16$\pi/c$). In the last panel of Fig. \ref{Fig3_cT}(b), we observe weak intensities corresponding to the $\beta$ band at 200 K presumably due to precursor effect associated to magnetic transitions often observed in various other systems \cite{bairo3,ca3co2o6}. The signature of $\kappa$ band disappears and $\delta$ band shifts towards $\epsilon_F$ (the separation between $\beta$ and $\delta$ band enhances) in the (0,0,$k_z$) spectra at higher temperatures. The $\delta$ band survives at 200 K and it is well discernible in the whole temperature range studied.

In order to capture the origin of these additional bands ($\delta$ and $\kappa$), we calculated the energy bands of CaFe$_2$As$_2$ in cT phase with lattice constants fixed to the experimental values at pressure, P = 0.63 GPa. The results for $z_{As}$ = 0.372 (value in tetragonal phase) are shown in Fig. \ref{Fig3_cT}(c). Two of the three Fe 3$d$ bands appear below $\epsilon_F$ and becomes degenerate at $\Gamma$ - they are marked with $\delta$. The third one, marked with $\kappa$, barely crosses $\epsilon_F$ close to the $\Gamma$ point and gives rise to a tiny hole pocket. This is demonstrated in the Fermi surface plot in Fig. \ref{Fig3_cT}(d), where the $\kappa$-band hole pocket survives at $\Gamma$ and exhibits weak $k_z$ dependence. If the experimentally observed value of $z_{As}$ (= 0.3663) at $P$ = 0.63 GPa is used in the calculations, which corresponds to a \textit{smaller pnictogen height} \cite{pnictogen-height}, even the hole pocket corresponding to $\kappa$-band moves below $\epsilon_F$ at $\Gamma$ point. Evidently, $z_{As}$ is a sensitive parameter for the survival/disappearance of the Fermi surfaces leading to charge fluctuations having an important implication in the physical properties of the system. The absence of hole pockets corresponding to $\delta$-band around $\Gamma$ also affects the FS nesting resulting into a loss of magnetic order although the Fe-moments are still finite - a good scenario for quantum spin fluctuations. The $\delta$ and $\kappa$ bands in the experimental results are strikingly similar to the theoretically observed energy bands of the Fermi surfaces corresponding to the cT phase possessing 3D topology.


From the above results, it is clear that although the magnetic \& structural transitions are quite similar in all the samples studied, the emergence of the $k_z$-dependence of the Fermi surface is sample dependent. We found that better quality samples possessing sharp transitions in the bulk properties along with bright, sharp \& well defined Laue spots in the $x$-ray Laue diffraction pattern, exhibits more distinct $\delta$ and $\kappa$ bands. Other samples exhibit these bands with lesser clarity presumably due to the disorder/distributed strain induced effects. A careful look at the dispersion of the $\kappa$ band in Fig. \ref{Fig3_cT}(b) (first panel) indicates that it does not cross $\epsilon_F$ as expected for the ambient phases, instead it moves towards higher binding energies; $\sim$25 meV in the 70 eV spectrum, which is $\Gamma(cT)$ point. Comparison of the experimental results and the calculated energy bands for different structural phases indicates that the $\delta$ and $\kappa$ bands are an unequivocal signature of the cT phase. In the band structure calculations, we found that the converged total energy for the antiferromagnetically ordered orthorhombic CaFe$_2$As$_2$ is the lowest indicating it as the ground state configuration. The converged energy for the nonmagnetic cT phase is close to the ground state energy; the calculations for all the other phases converged to higher total energies \cite{ctdft_khadiza}. Thus, a small perturbation/strain due to the application of pressure and/or preparation condition is expected to influence the crystal structure significantly and bring the system to cT phase. Thermal compression would enhance the strain in the samples, which explains the shift of the $\delta$ and $\kappa$ bands on cooling.

Finite electron-electron Coulomb repulsion strength among Fe 3$d$ electrons leads to band narrowing and local character of the electrons that enhances spin fluctuation in addition to its influence on charge excitations. On the other hand, strong Hund's coupling among the 3$d$ electrons tries to align their spin moments without much influence on charge excitations and enhances spin fluctuations in the system. These competing interactions leads to significant reduction of Fe-moment from its atomic value and the ground state is a spin density wave (SDW) state with Fe atoms possessing magnetic moment close to a Bohr magneton \cite{mag_CaFe2As2,mag_BaFe2As2,mag_LaoFeAs}. Superconductivity is realized in these systems via suppression of the magnetic order, which is believed to be due to the spin fluctuations \cite{spinfluc_wang,spinfluc_allan}. It is evident here that the intrinsic strain brings these materials towards quantum fluctuations; the proximity to vanishing hole pockets around $\Gamma$-point due to the cT phase leads to charge fluctuations as well as spin fluctuations due to the proximity of vanishing Fermi surface nesting.


The Fe-based compounds are complex presumably due to the presence of phases hidden to various experimental probes and the electronic properties are often found to be puzzling. We discover signature of cT phase hidden in the normal structural phase employing ARPES, which became possible due to the detailed study of the Fermi surfaces at intermediate temperatures and employment of high photon energies. The coexistence of the cT phase within the ambient phases enhances the quantum fluctuations in the system and may be responsible for various exotic electronic properties in these materials.


K.M., K.A. and S.P. thank ICTP for providing financial support for the measurements at Elettra, Trieste. K. M. acknowledges financial assistance from the Department of Science and Technology, Govt. of India under J.C. Bose Fellowship program and Department of Atomic Energy, Govt. of India under DAE-SRC-OI award scheme.

\section*{AUTHOR CONTRIBUTIONS}
K.A., G.A., S.T., S.P, S.M. and K.M. carried out the ARPES measurements. A.T. provided the sample. P.M. and P.S. helped during the measurements at the VUV beamline, Elettra. K. A. did the band structure calculations. K.A. \& K.M. analysed the ARPES data. K.M. proposed the project, supervised the work and wrote the manuscript. All the authors read the manuscript and provided their input.



\end{document}